\documentclass[11pt]{article}
\usepackage{fullpage}
\usepackage{authblk}
\usepackage{graphicx}
\usepackage{amsmath}
\usepackage{amssymb}
\usepackage{multirow}
\usepackage{url}
\usepackage{xcolor}
\usepackage{listings}

\begin{document}

\title{Accelerating Lattice QCD Multigrid on GPUs Using Fine-Grained Parallelization}

\author[1]{M. A. Clark \thanks{mclark@nvidia.com}}
\author[2]{B\'alint Jo\'o \thanks{bjoo@jlab.org}}
\author[3]{Alexei Strelchenko \thanks{astrel@fnal.gov}}
\author[4]{Michael Cheng \thanks{mchengcit@gmail.com}}
\author[5]{Arjun Gambhir \thanks{asgambhir@email.wm.edu}}
\author[6]{Richard C. Brower \thanks{brower@bu.edu}}

\affil[1]{NVIDIA Corporation, 2701 San Tomas Expressway, Santa Clara, CA 91214, USA}
\affil[2]{Thomas Jefferson National Accelerator Facility (Jefferson Lab), Newport News, VA 23606, USA}
\affil[3]{Fermi National Accelerator Laboratory, Batavia, IL 60510-5011, USA}
\affil[4]{Center for Computational Science, Boston University, Boston, MA 02215, USA}
\affil[5]{The College of William and Mary, Williamsburg, VA 23187-8795, USA}
\affil[6]{Physics Department, Boston University, Boston, MA 02215, USA}

\setcounter{Maxaffil}{0}
\renewcommand\Affilfont{\itshape\small}

\providecommand{\preprint}[1]{\textbf{Preprint Numbers: } #1}

\maketitle

\begin{abstract}
  The past decade has witnessed a dramatic acceleration of lattice
  quantum chromodynamics calculations in nuclear and particle
  physics. This has been due to both significant progress in
  accelerating the iterative linear solvers using multi-grid
  algorithms, and due to the throughput improvements brought by
  GPUs. Deploying hierarchical algorithms optimally on GPUs is
  non-trivial owing to the lack of parallelism on the coarse grids,
  and as such, these advances have not proved multiplicative. Using
  the QUDA library, we demonstrate that by exposing all sources of
  parallelism that the underlying stencil problem possesses, and
  through appropriate mapping of this parallelism to the GPU
  architecture, we can achieve high efficiency even for the coarsest
  of grids. Results are presented for the Wilson-Clover
  discretization, where we demonstrate up to 10x speedup over present
  state-of-the-art GPU-accelerated methods on Titan. Finally, we look
  to the future, and consider the software implications of our
  findings.
\end{abstract}

\section{Introduction}
\vspace{-1mm}

It is well known that continued advances in high performance computing
(HPC) have been powered by ever increasing transistor density with each
generation, i.e., Moore's Law.  While these advances have continued unabated
over the course of 40 years, over the past decade, due to the loss of
Dennard scaling, their manifestation has evolved from faster into more
parallel.  This evolution has had a significant impact on software,
since software has had to evolve to enable more parallel processors to
be utilized.

Simultaneously, a less-well quantified advance has occurred, that
being the one of algorithmic advances: super-linear acceleration of
fundamental algorithmic building blocks that underpin computational
science.  Examples abound from the various disciplines of
computational science, e.g., multigrid (MG)~\cite{10.2307/2006422}, Fast Multipole
Method~\cite{ROKHLIN1985187}, Strassen's matrix multiplication~\cite{strassen}, etc.
Taken together, the combination of algorithmic and machine advances have
revolutionized computational science.  Calculations, thought impossible
30 years ago, are now routine.  However, this meteoric gain in science
throughput is only possible if this combination is {\it
  multiplicative}.  Sadly, the recent trends in microprocessor
evolution have been divergent with respect to algorithmic advances:
the former is becoming ever more parallel, and requires ever more
locality (minimization of data movement); in the case of the latter,
the critical component of exponential algorithm acceleration is often
a serial component, and/or requires non-local communication.

In the present work, we consider the case of lattice quantum
chromodynamics (LQCD), a numerically feasible formulation of the theory
of the strong force that underpins nuclear and particle physics.  This
grand challenge application is extremely computationally demanding,
often consuming 10\%-20\% of public supercomputing cycles around the world.
While the field has existed for around forty years, it is only
recently that both machines and algorithms have advanced enough
to bring LQCD predictions to the sub-1\% error level, e.g., LQCD calculations
are now finally capable of making high-precision predictions for
comparison against large-scale accelerator facilities, such as the
Large Hadron Collider at CERN. 

Graphics processing units (GPUs) have proved to be
both a popular and efficient platform on which to build and deploy HPC
applications.  Present GPUs typically feature thousands of floating point units
coupled to a very wide and fast memory bus.  GPUs are programmed using
a threaded model, utilizing thread oversubscription to hide latency,
and require upwards of ten thousand active threads in order to
saturate their performance.  GPUs are very well
suited to LQCD, since these computations feature a lot of trivial data
parallelism, as well as having highly-regular memory accesses, which
lead to high bandwidth utilization.

One of the largest algorithmic advances in recent years has been the
removal of the critical slowing down in the iterative linear solvers
that has plagued LQCD computations: as the quark mass parameter is
reduced, the condition number of the quark-gluon interaction matrix
(known as the Dirac matrix), which must be solved in most LQCD
calculations, diverges.  This difficulty can be almost completely
removed through the use of hierarchical preconditioners such as
adaptive MG~\cite{Babich:2010qb}.  The end result is an {\em
  algorithmic} acceleration of the linear solver by potentially
upwards of \(10\times\).
This improved algorithmic scaling will enable larger and more precise
computations in the future.

Thus it is obvious to seek the full multiplicative improvement from
algorithm and machine.  However, GPUs represent an extremely
challenging architecture on which to deploy an efficient MG
algorithm: the coarse-grid computations that underpin an efficient
algorithm are by definition increasingly serial workloads.  In this
work, using the QUDA library, we demonstrate that through identifying
and exposing all of the underlying latent parallelism in the
coarse-grid computations we can create a highly efficient MG
algorithm implementation.  From an LQCD-workload perspective, we focus on the
throughput-based workloads of LQCD, that exhibit a lot of task
parallelism, and seek to optimize total job throughput.  The end
result is that we are able to accelerate the analysis workloads by up
to \(10\times\)
over present state-of-the-art LQCD computations on GPUs.  In the
future, we shall refocus on the other critical workflow in LQCD, {\it
  gauge generation}, which
typically uses a Markov-chain Monte Carlo algorithm.  In this stage of
computation there exists only data parallelism with limited task
parallelism, so strong scaling a given problem set is critical for
increasing science throughput.

This work is cross cutting since hierarchical algorithms are important
for many computational-science disciplines.  Moreover, while a given
discipline may not be parallelism challenged on current-generation
many-core processors, as we trend towards the exascale and beyond,
this will be increasingly the case for many fields.  Furthermore, we
cannot rely on increasing problem set sizes to meet parallelism
requirements since in many cases the computational cost grows
super-linearly with problem size, so parallelism must be sought
elsewhere.

This paper is organized as follows: in \(\S\)\ref{sec:previous} we highlight
previous work in this area, we give an overview of the LQCD
computations we seek to accelerate in \(\S\)\ref{sec:lqcd}, and in
\(\S\)\ref{sec:quda} introduce the QUDA library that underpins this work.
We give an overview of our hierarchical framework in
\(\S\)\ref{sec:hier} and discuss fine-grained parallelization in
\(\S\)\ref{sec:fine}.  We show strong-scaling performance curves from the
Titan supercomputer in \(\S\)\ref{sec:results}; we discuss the software implications
in \(\S\)\ref{sec:software} and future research topics in
\S\ref{sec:future-work} before finally concluding with
\(\S\)\ref{sec:conclusion}.

\vspace{-1mm}
\section{Previous Work}
\vspace{-1mm}
\label{sec:previous}

Lattice QCD calculations on GPUs were originally reported in
\cite{Egri2007631} where the immaturity of using GPUs for general
purpose computation necessitated the use of graphics APIs.  Since the
advent of CUDA in 2007, there has been rapid uptake by the LQCD
community.  Notable
publications include multi-GPU parallelization
\cite{Babich:2010:PQL:1884643.1884695, Shi:2011ipdps,
  Alexandru:2011sc}, use of additive Schwarz preconditioning to
improve strong scaling \cite{Babich:2011np}, software-managed
cache-blocking strategies \cite{inpar2012}, and JIT-compilation to enable the
offload of the entire underlying data-parallel framework of the Chroma \cite{Edwards:2004sx} code without any top-level changes \cite{Winter:2014dka}.  This work concerns the QUDA library~\cite{Clark:2009wm}, of which we give an overview in \S\ref{sec:quda}.

The basic principle of gauge-invariant MG for LQCD was begun
with Projective MG in the early
1990s~\cite{Brower:1991xv}. However only with the advent of an
adaptive MG algorithm introduced recently into LQCD
\cite{Brannick:2007ue, Babich:2010qb} was critical slowing down removed
completely in the limit of zero quark mass. In parallel, the related
approach of {\it Inexact Deflation} was introduced~\cite{Luscher:2007se}; the essential
difference between these approaches being that the former utilizes a
multiplicative coarse-grid correction, while the latter uses an additive
correction.  Evolutions of the algorithm introduced include combining
it with red-black (or even-odd in LQCD parlance) preconditioning
\cite{Osborn:2010mb} and improving the strong scaling through
utilizing a Schwarz-preconditioned domain-decomposition smoother to
reduce inter-node communication \cite{Frommer:2013fsa,
  Frommer:2013kla}.  These publications have all utilized more
traditional CPU-based clusters. Implementations for Intel Xeon Phi
Knights Corner have been reported in
\cite{Heybrock:2015kpy,Richtmann:2016kcq}.  To the best of our
knowledge there have been no publications concerning the efficient deployment of
these MG algorithms for LQCD on GPUs.

The successful deployment of geometric MG on GPUs has been shown
in porting the HPGMG benchmark to GPUs~\cite{hpgmg-gpu}.

\vspace{-1mm}
\section{Lattice Quantum Chromodynamics}
\vspace{-1mm}
\label{sec:lqcd}

Weakly coupled field theories such as quantum electrodynamics
can by handled with perturbation theory.
In QCD, however, at low
energies perturbative expansions fail and a non-perturbative 
method is required. LQCD is the only known, model
independent, non-perturbative tool currently available to perform
QCD calculations.

LQCD calculations are typically Monte-Carlo evaluations of  
Euclidean-time path integrals.  A sequence of configurations of the gauge fields
is generated in a process known as {\em
  configuration generation}. The gauge configurations are
importance-sampled with respect to the lattice action and
represent a snapshot of the QCD vacuum.  Configuration generation is
inherently sequential as one configuration is generated from the
previous one using a stochastic evolution process. As the lattice structure admits data parallelism, many variables can
be updated simultaneously, and the focused power of capability computing
systems is essential to carry out this phase of the calculations.  
Once the field configurations have been generated, one moves on to the second stage of the calculation, known as {\em analysis}. In this phase, observables of interest are
evaluated on the gauge configurations in the ensemble, and the results
are then averaged appropriately, to form {\em ensemble averaged}
quantities. It is from the latter that physical results such as
particle energy spectra can be extracted.  The analysis phase can be
task parallelized over the available configurations in an ensemble and
is thus extremely suitable for capacity-level work on clusters, and large
ensemble calculations can also make highly effective use of capability-sized 
partitions of leadership  supercomputers.  It is the analysis phase that
we focus upon in this work.

\vspace{-1mm}
\subsection{Dirac PDE discretization}
\vspace{-1mm}

The fundamental interactions of QCD, those taking place between quarks
and gluons, are encoded in the quark-gluon interaction differential
operator known as the Dirac operator.  As is common in PDE solvers,
the derivatives are replaced by finite differences.  Thus on the
lattice, the Dirac operator becomes a large sparse matrix, \(M\),
and the calculation of quark physics is essentially reduced to many
solutions of systems of linear equations given by
\vspace{-1mm}
\begin{equation}
Mx = b.
\label{eq:linear}
\end{equation}
Computationally, the brunt of the numerical work in LQCD for both the
gauge generation and analysis phases involves solving such linear
systems.

A small handful of discretizations of the continuum operator are in common use, differing in
their theoretical properties.  Here we focus on the widely-used Sheikholeslami-Wohlert \cite{Sheikholeslami:1985ij} form for $M$,  
colloquially known as the {\em Wilson-Clover}, or simply just {\em Clover} discretization.

\vspace{-1mm}
\subsection{Wilson-Clover matrix}
\vspace{-1mm}
 The Wilson-Clover matrix is a central-difference discretization of the
Dirac operator, with the addition of a diagonally-scaled Laplacian to
remove the infamous fermion doublers (which arise due to the red-black
instability of the central-difference approximation).  When acting in
a vector space that is the tensor product of a four-dimensional
discretized Euclidean spacetime, {\it spin} space, and {\it color}
space it is given by
\vspace{-1mm}
\begin{align}
 M_{x,x'} &= - \frac{1}{2} \displaystyle \sum_{\mu=1}^{4} \bigl(
 P^{-\mu} \otimes U_x^\mu\, \delta_{x+\hat\mu,x'}\, + P^{+\mu} \otimes
 U_{x-\hat\mu}^{\mu \dagger}\, \delta_{x-\hat\mu,x'}\bigr) \nonumber \\
&\quad\, + (4 + m + A_x)\delta_{x,x'}
\label{eq:Mclover}
\end{align} 
 Here \(\delta_{x,y}\) is the Kronecker delta; \(P^{\pm\mu}\) are
\(4\times 4\) matrix projectors in {\it spin} space; \(U\) is the QCD
gauge field which is a field of special unitary $3\times 3$ (i.e.,
SU(3)) matrices acting in {\it color} space that are ascribed to the links between the
spacetime sites (and hence are referred to as {\it link matrices}); \(A_x\)
is the \(12\times12\) Clover matrix field acting in both spin and
color space,

corresponding to a first-order
discretization correction; and \(m\) is the quark-mass parameter.  The
indices \(x\) and \(x'\) are spacetime indices (the spin and color
indices have been suppressed for brevity).  This matrix acts on a
vector consisting of a complex-valued 12-component \emph{color-spinor}
(or just {\em spinor}) for each point in spacetime.  We refer to the
complete lattice vector as a spinor field.

\begin{figure}[htb]
\begin{center}
\includegraphics[trim={0 0.4cm 0 0.4cm},clip,width=2.4in]{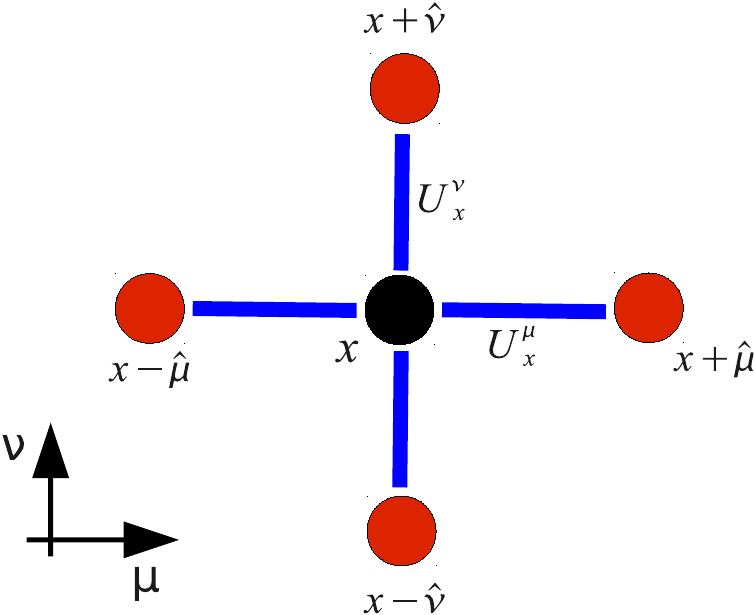}
\end{center}
\caption{\label{fig:dslash}The nearest-neighbor stencil part of the
 lattice Dirac operator, as defined in (\ref{eq:Mclover}), in the $\mu-\nu$
 plane.  The \emph{color-spinor} fields are located on the
 sites. The SU(3) color matrices $U^\mu_x$ are associated with the links.}
\end{figure}

\vspace{-1mm}
\subsection{Krylov solvers}
\vspace{-1mm}
 Iterative Krylov
solvers are typically used to obtain solutions to Equation (\ref{eq:linear}),
requiring many repeated evaluations of the sparse matrix-vector
product.  The Wilson-Clover matrix is non-Hermitian, so either
Conjugate Gradients \cite{Hestenes:1952} on the normal equations (CGNE
or CGNR) is used, or more commonly, the system is solved directly
using a non-symmetric method, e.g., BiCGStab \cite{vanDerVorst:1992}.
Red-black preconditioning is almost always
used to accelerate the solution finding process for this system, where
the nearest-neighbor property of the \(M\) matrix is exploited
to solve the Schur complement system~\cite{Degrand1990211}. 

The quark mass controls the condition number of the matrix, and hence
the convergence of such iterative solvers.  Unfortunately, physical
quark masses correspond to highly singular matrices.  Given that
current lattice volumes are up to \(10^{10}\)
degrees of freedom, this represents an extremely
computationally demanding task.  For virtually all LQCD computations,
the linear solver accounts for 80--99\% of the execution time.
Without resorting to eigenvalue deflation or hierarchical approaches,
the use of mixed-precision BiCGStab combined with red-black
preconditioning represents the state of the art.

\vspace{-1mm}
\subsection{Adaptive Geometric Multigrid}
\vspace{-1mm}

The problem of critical slowing down with the quark mass was long a
source of consternation with LQCD computations.  While the problem can
be alleviated with eigenvector-deflation algorithms, these algorithms
scale quadratically with the volume owing to the spectral density
scaling approximately linearly with volume, as well as the number of
operations scaling linearly with volume.  The {\it scalable} solution
(with respect to problem size) to critical slowing down is the use of
adaptive MG.

The scope of this paper is too narrow to permit a careful treatment of
the adaptive geometric algorithm deployed for LQCD, so we focus on
describing the key parts of the algorithm.  As with standard
MG, the two critical ingredients to a convergent algorithm are:
\begin{itemize}
\item A smoother that effectively reduces high-frequency
error components.
\item Prolongation and restriction operators that define a coarse grid
  that captures the near null-space components of the fine-grid linear
  operator.
\end{itemize}
What complicates the LQCD case is that the near-null space modes of
the Dirac operator (equation \ref{eq:Mclover}) are highly oscillatory
due to the underlying dependence on the stochastic gauge field, and
are not captured by a conventional geometric MG method.
The coefficients of the prolongator must therefore be set adaptively
to capture this null space by using vectors that are rich in the
low-frequency error components of the Dirac operator.  Critical to the
success of this method is the {\it weak-approximation} property of the
null space: while eigenvectors are globally
(bi-)orthogonal, locally they are co-linear; meaning that a small
collection of near-null-space vectors, when partitioned into disjoint
local subsets (or {\it aggregates}), form a basis that spans the
majority of the near-kernel subspace of the
operator~\cite{Brezina2004}.  Given that the problem is discretized on
a homogeneous hypercube, there is no need to use algebraic
aggregation, and thus the shape of the aggregates are regular
non-overlapping hypercubes.  Hence the algorithm is both adaptive {\it
  and} geometric.

The adaptive geometric MG setup is as follows
\begin{enumerate}
\item Iterate the homogenous system \(Mx=0\) with a random initial
  guess \(x_0\).  After \(k\) iterations the resulting
  error vector \(e_k = -x_k\) will be rich in slow-to-converge
  eigenmodes of the operator \(M\).
\item Repeat 1) until we obtain enough candidate vectors to capture
  the near-null space of the operator \(M\).
\item Partition the set of null-space vectors \(v\) into disjoint
  aggregates, and perform a QR decomposition on these blocks to form a
  local orthonormal basis.  This forms the columns of the prolongation operator \(P\).
\item Compute the resulting coarse-grid operator from the Galerkin
  product \(\hat{M} = P^{\dagger} M P\).
\end{enumerate}

By virtue of the fact that the original fine-grid operator is nearest
neighbor only, the coarse-grid operator retains this property.
However, due to the aggregation summing over spin and color components of the null-space
vectors, the simple tensor-product structure  between spin
and color spaces on the fine grid is lost on the coarse grid. This 
manifests itself as a loss in sparsity: the link
matrices that exist between lattice sites are of size
\(\hat{N}_s\hat{N}_c\times \hat{N}_s\hat{N}_c\),
where \(\hat{N}_c\)
is the number of null-space vectors used to represent the fine-grid
null space; this corresponds to the effective number of colors on the coarse
grid; \(\hat{N}_s=2\)
is the number of spin degrees of freedom on the coarse
grid.\footnote{The upper and
  lower spin components are aggregated separately, resulting in
  \(\hat{N}_s=2\): this {\it chirality} preservation
  ensures that a vector rich in right low modes is also rich in
  left low modes, and allows us to define the restrictor as the
  Hermitian conjugate of the
  prolongator~\cite{Brower:1991xv,Babich:2010qb}.}
The resulting coarse-grid operator takes the form,
\begin{equation}
  \hat{M}_{x,x'} = -\sum_{\mu}^{4} \left[Y^{-\mu}_{x}\delta_{x+\hat{\mu},x'} + Y^{+\mu\dagger}_{x-\hat{\mu}} \delta_{x-\hat{\mu},x'}\right]                                                                
-\delta_{x,x'},
\label{eq:coarse-dslash}
\end{equation}
where the matrix field \(Y\) is the coarse link field.

Aggregates of size \(2^4-8^4\)
are usually chosen, with 20-30 vectors required to capture enough of
the null-space.  The algorithm is recursive, and the setup process can
be repeated for an arbitrary number of levels; the coarse-grid
operator retains the form in Equation \ref{eq:coarse-dslash} upon
subsequent coarsening.

\vspace{-1mm}
\section{The QUDA Library}
\vspace{-1mm}
\label{sec:quda}

QUDA (QCD on CUDA) is a library that aims to accelerate LQCD
computations through offload of the most time-consuming components of
an LQCD application to NVIDIA GPUs.  It is a package of optimized CUDA C++
kernels and wrapper code, providing a variety of optimized linear
solvers, as well as other performance critical routines required for
LQCD calculations.  All algorithms can be run distributed on a cluster
of GPUs, using MPI to facilitate inter-GPU communication.  It has been
designed to be easy to interface to existing code bases, and in this
work we exploit this interface to use the popular LQCD application
Chroma~\cite{Edwards:2004sx}.  The QUDA library has attracted a
diverse developer community and is being used in production at U.S.\
national laboratories, as well as in locations in Europe and India.
The latest development version is always available in a
publically-accessible source code repository~\cite{githubQUDA}.

The general strategy is to assign a single GPU thread to each lattice
site. Each thread is then responsible for all memory traffic and
operations required to update that site on the lattice given the
stencil operator.  Since the computation always takes place on a
regular grid, a {\it matrix-free} approach is always used, with the
thread coordinate index computed dynamically (see for example Listing
\ref{listing:index}). Maximum memory bandwidth is obtained by
reordering the spinor and gauge fields to achieve memory coalescing,
e.g., using structures of float2 or float4 arrays, and using the
texture cache where appropriate.  Memory-traffic reduction is employed
where possible to overcome the relatively low arithmetic intensity of
the Dirac matrix-vector operations, which would otherwise limit
performance.  Strategies include: (a) using compression for the
\(SU(3)\) gauge matrices to reduce the 18 real numbers to 12 (or 8) real numbers
at the expense of extra computation; (b) using similarity transforms
to increase the sparsity of the Dirac matrices; (c) utilizing a custom
16-bit fixed-point storage format (hereon referred to as half
precision) together with mixed-precision linear solvers to achieve
high speed with no loss in accuracy~\cite{Clark:2009wm}.

The library has been designed to allow for maximum flexibility with
respect to algorithm parameter space and maximum performance.  For
example, all lattice objects (fields) maintain their own precision and
data ordering as a dynamic variable.  This allows for run-time policy
tuning of algorithms; these parameters are then bound at kernel launch
time when the appropriate C++ template is instantiated corresponding
to these parameters.  While this flexibility does provide maximum
performance it does come at the expense of library binary size.

Key to achieving high performance is the use of auto-tuning: all kernel
launch parameters (block size, grid size, shared memory size) are
auto-tuned upon first call and cached for subsequent reuse.  The autotuner can also tune for
arbitrary algorithm policy choices outside of kernel launch parameters
(see \S\ref{sec:fine-perf}).

\vspace{-1mm}
\section{Heterogeneous Software Architecture}
\vspace{-1mm}
\label{sec:hier}

MG has both highly parallel throughput-limited parts (fine
grids) and more serial, latency-limited parts (coarse grids).  Simultaneously, we have an architecture that is composed of throughput (GPUs)
and latency optimized (CPUs) processors, so it is natural to
consider whether mapping the coarse grids onto the CPU is more
efficient.  Thus, from the outset, the software-design goal was to enable QUDA to
run on the complete {\it heterogeneous} architecture in the
most efficient manner.  To this end, much of QUDA was redesigned to
abstract the algorithm from the underlying architecture-specific
implementation.  Algorithms are expressed in terms of generic fields,
with no indication at the algorithmic level as to where the
computation is to be executed.  Similar to the data order or precision, the
{\it location} of the data is elevated to a run-time object member,
and when executing an algorithm on an object, the object's
location is queried, and the architecture-specific code is executed.

While abstracting the execution location from the algorithm makes it
simple to write architecture agnostic high-level algorithms, this
still leaves the problem of having to maintain two copies of each
computational routine.  To remove this burden, we utilize the fact
that CUDA and CPU kernels can call common functions if those functions
have the \texttt{\_\_device\_\_ \_\_host\_\_} decorators applied to
their declaration.  In this way we can use a common codepath for the
bulk of the computational routines: the GPU kernel declarations being
nothing more than stubs that set indices based on the thread id and
call the common function with this index parameter; the CPU functions
similarly are stubs that contains {\em for} loops over the index range (with
optional OpenMP parallelization) that call the
common function.  The data order is abstracted into generic accessor
functors, this allows for different data ordering for the CPU and GPU
variants, allowing for optimal deployment on both platforms.

\lstset{language=C++,caption={Example of how an axpy computation is
    deployed using a single code path for GPU and CPU.},label=listing:arch}
\begin{lstlisting} 
template <typename Arg>
__device__ __host__ void axpy(Arg &arg, int i)
{ arg.y(i) += arg.a * arg.x(i); }

template<typename Arg> // GPU code path 
__global__ void axpyGPU(Arg arg) {
  int i = blockDim.x*blockIdx.x + threadIdx.x;
  if (i >= arg.length) return;
  axpy(arg,i);
}

template<typename Arg> // CPU code path 
void axpyCPU(Arg &arg) {
#pragma omp for
  for (int i=0; i<arg.length; i++) axpy(arg,i);
}
\end{lstlisting}

Within the context of MG, it is only at the inter-grid
interface, i.e., prolongation and restriction, where it makes sense to
switch from one architecture to the other since within a grid we have
a roughly constant degree of parallelism and workload.  Take for
example the restrictor, where we have as input a fine-grid field on
the GPU, and output a coarse-grid field on the CPU.  Given that the
data input is far greater than the output, and the relative narrow
PCIe connection between GPU and CPU, we compute the restriction on the
GPU and copy the result to the CPU.  The converse is true for
prolongation.

The consideration of when to offload work to the CPU versus doing it
on the GPU is one that has no simple answer, since it depends on many
factors, including
\begin{itemize}
\item Algorithmic parameters: precision, problem size, etc.
\item Relative performance of GPUs and CPUs and the ratio of their numbers per node.
\item The CPU-GPU connection bandwidth
\item Whether parallel work can be overlapped during the execution of
  this phase
\item Network latency differences between the two memory spaces
\end{itemize}

The optimal strategy on one machine will thus be different from
another.  These topics are left unexplored in this work, since as
shall be shown below we achieve excellent performance maintaining the
entire calculation on the GPU.  Given that the QUDA hierarchical
framework has been designed to facilitate the offload of arbitrary
computation to the CPU as a {\it run-time policy} decision, in the
future we envisage extending the QUDA autotuning framework to tune for
computation location as a policy decision.

\vspace{-1mm}
\section{Fine-grained Parallelization}
\vspace{-1mm}
\label{sec:fine}

Prior to this work, all algorithms in QUDA have utilized only the
trivial data parallelism over grid points: each thread corresponds to
an output grid point, which applies the stencil as a gather operation.
This mapping works well for large grids, where large means we
have enough grid points, and thus threads, to hide the GPU instruction
and memory latencies.  Typically this means O(10,000) grid points or
greater.  In other application domains, coarser-grained parallelizations are possible
and may prove optimal if there is sufficient instruction-level
parallelism (ILP) to hide
latency~\cite{Micikevicius:2009:FDC:1513895.1513905, volkov-stencil}.

For our hierarchical problem, where the coarsest grid size will have
as few as \(2^4\)
grid points, or 16-way grid parallelism, such an approach will not be
suitable for a GPU.  Thus it is critical that we expose all sources of
parallelism available to us, and ascertain how to optimally map these
to the architecture.

Increasing the degree of parallelization is also important for load
balancing.  We seek an algorithm that is scalable to an arbitrary
number of processing cores, and do not want to rely on a specific core
count or vector width.  Thus, it is critical that we have a sea of
lightweight threads that scale without load balancing or edge effects
being apparent.

In the analysis that follows we focus on the coarse-grid operator.
Similar analysis applies to the prolongator and restrictor, the
specifics of these are covered in \S\ref{sec:fine-transfer}.

\vspace{-1mm}
\subsection{Grid Parallelization}
\vspace{-1mm}

As noted previously, grid parallelism is the trivial data parallelism
that arises from assigning one thread per lattice site.  There are no
inter-thread dependencies, and so the parallelization is trivial.  The
one dimensional global thread index is trivially mapped to the
lattice-site index as shown below.  This necessarily
involves integer division which adds a non-trivial overhead, however, this is amortized 
in general when using a coarse-grained decomposition.
\lstset{language=C++,caption={Example of how the thread index maps to
    the lattice coordinates},label=listing:index}
\begin{lstlisting}
// X[] holds the local lattice dimension  
int idx = blockIdx.x*blockDim.x + threadIdx.x;
int tmp1 = idx / X[0];
int tmp2 = tmp1 / X[1]; 
x[0] = idx - tmp1 * X[0];
x[1] = tmp1 - tmp2 * X[1];
x[3] = tmp2 / X[2];
x[2] = tmp2 - x[3] * X[2];
// x[] now holds the thread coordinates
\end{lstlisting}

\vspace{-3mm}
\subsection{Color and Spin Parallelization}
\vspace{-1mm}

If we consider Equation \ref{eq:coarse-dslash}, the output per site is
a color-spinor vector.  Each element of this vector is independent,
and so like the grid points we can trivially split the computation
between threads.  Thus we potentially have up to
\(\hat{N}_s\times \hat{N}_c\)-way
additional parallelism that can be exploited.  In general we have
found using the maximal fine-grained decomposition to be optimal, with
each thread computing a single dot product of the matrix-vector
multiplication (i.e., one color and spin per thread).  We utilize the
y-dimension of the CUDA thread index to map to which output element we
are computing.  In order to optimize between the temporal locality
over the y-thread dimension (the input vector is common between
threads) versus the need to spread the computation in order to improve
load balancing across the GPU the block size in the y dimension is
autotuned.  \lstset{language=C++,caption={Example of how the y-thread
    index maps to resulting color-spin
    indices},label=listing:color-spin}
\begin{lstlisting}
// Nc is the total number of colors
// Mc is the number of colors per thread
int yIdx = blockDim.y*blockIdx.y + threadIdx.y; 
int s = yIdx / (Nc/Mc);
int color_subset = (yIdx % (Nc/Mc)) * Mc;
\end{lstlisting}

\vspace{-3mm}
\subsection{Stencil Direction Parallelization}
\vspace{-1mm}

The next level of parallelism we can exploit is that provided by the
stencil direction: the \(\mu\)
index from Equation \ref{eq:coarse-dslash} as well as the forwards and
backwards directions.  This provides up to an additional eight-way
parallelism, however, we must sum the partial results before writing
to main memory.  In CUDA this is easily done by
ensuring that all threads responsible for a given output element are
assigned with the same thread block:
\begin{enumerate}
\item Each thread computes its contribution to the sum.
\item Each thread writes its result to shared memory.
\item Synchronize the thread block.
\item Threads responsible for, say,  \(\mu=0\) and forwards
  direction, sum all results from shared memory and writes to global memory.
\end{enumerate}

Parallelization over the stencil direction is mapped to the z-thread
dimension, with the mapping \(\mu*2 + dir\) being a trivial variation of Listing \ref{listing:color-spin}.  On
larger grids it was found to be detrimental to parallelize the stencil
direction, and the optimal degree of splitting varies across GPU
architecture.  Thus, the degree to which the stencil is parallelized
is a templated parameter, and the autotuner ascertains the optimal
splitting for a given problem size and architecture.

\vspace{-1mm}
\subsection{Dot Product Parallelization}
\vspace{-1mm}
\label{sec:fine-rows}

Since the grid sites are assigned the fastest varying thread index (x
dimension), on the smallest grids, e.g., \(V=2^4\),
we will not have enough grid sites to fill a warp\footnote{A warp
  corresponds to 32 consecutive threads in CUDA, and
  is equivalent to the SIMD vector length.} and guarantee full
SIMD efficiency, since warp threads could diverge.
To remedy this, we can partition the dot-product computation that
corresponds to multiplying a row of the link-matrix by a site vector.
In general, such dot products are unsuitable for parallelizing since
they are tightly coupled and combining the results from individual dot
products would require costly synchronization.  However, here we only
seek to expose enough parallelism to enable us to fill the warp, e.g.,
2-way splitting of the dot product for 16 grid points.  To do so, we
utilize the warp-shuffle instruction that allows for low-latency
communication between threads in the same warp without the need to
synchronize.\footnote{Generics~\cite{generics} is used to
  provide warp-shuffle for generic types.}

\lstset{language=C++,caption={Example \(N\)-way
    dot product partitioning using the warp shuffle instruction to
    maximize thread efficiency},label=listing:shuffle}
\begin{lstlisting}
const int warp_size = 32; // warp size
const int n_split = 4; // four-way warp split
const int grid_pts = warp_size / n_split;
complex<real> sum = 0.0;
for (int i=0; i<N; i+=n_split)
  sum += a[i] * b[i];

// cascading reduction
for (int offset = warp_size/2; offset >= grid_pts; offset /= 2)
  sum += __shfl_down(sum, offset);
// first grid_pts threads now hold result
\end{lstlisting}

Finally, we expose ILP, allowing for intra-thread latency hiding.
This optimization is more important for the Kepler architecture that
Titan features, since it has higher dependent instruction latency
(nine clock cycles) than the more recent Maxwell and Pascal (six clock cycles).

\lstset{language=C++,caption={Example of how ILP is exploited to
    increase throughput for an \(N\)-way complex dot-product type computation},label=listing:ilp}
\begin{lstlisting}
const int n_ilp = 2; // two-way ILP
complex<real> sum[n_ilp] = { };
for (int i=0; i<N; i+=n_ilp)
  for (int j=0; j<n_ilp; j++)
    sum[j] += a[i+j] * b[i+j];

complex<real> total = static_cast<real>(0.0);
for (int j=0; j<n_ilp; j++) total += sum[j];
\end{lstlisting}

\subsection{Performance}
\label{sec:fine-perf}

\begin{figure}[htb]
\begin{center}
\includegraphics[trim={0.06cm 0.08cm 0.05cm 0.1cm},clip, width=3.3in]{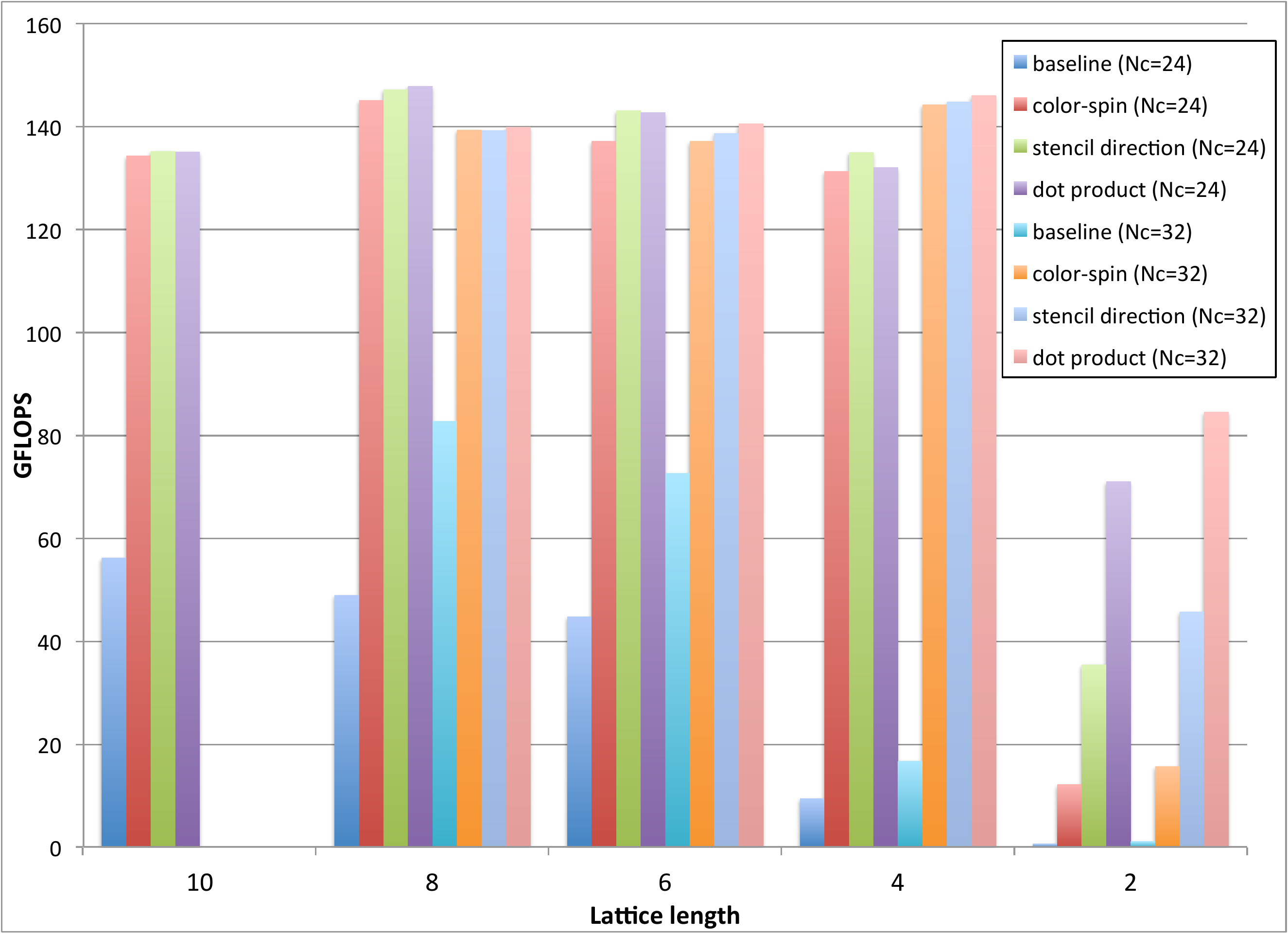}
\end{center}\vspace{-2mm}
\caption{\label{fig:coarse-dslash-perf}Single-precision performance of the
  coarse-grid operator as a function of decreasing lattice size for 24
  and 32 colors (Tesla K20X, CUDA 7.5, GCC 4.9).}
\end{figure}

In Figure \ref{fig:coarse-dslash-perf} we show the performance of the
coarse operator as a function of decreasing lattice size, with each
strategy being cumulative with previous ones.  The Tesla K20X used
here matches that of the Titan used in our final result in
\S\ref{sec:results}.  Given that the arithmetic intensity (in FP32) of
the coarse operator is close to unity, 140 GFLOPS represents around
80\% of achievable STREAM bandwidth, and so represents a reasonable
performance upper bound.  In comparison, the Wilson-Clover operator
sustains \(\sim 400\) GFLOPS
on equivalent fine-grid size.  The former is much smaller than the
latter due to the decreased temporal locality arising from the loss in
tensor structure.  For all but the smallest lattice size, the addition
of color-spin parallelization is enough to saturate performance.  On
the \(2^4\)
lattice, all sources of parallelism are necessary, and even then
performance is not saturated.  Profiling revealed this to be due to
the fixed indexing cost that represents the Amdahl's law limiter,
e.g., Listing \ref{listing:index}.  We note that on the \(2^4\)
lattice with 32 colors, the fine-grained parallelization results in
32768-way parallelism, instead of the na\"{i}ve 16-way parallelism,
and resulted in a 100x speedup.  Future optimization in this area
could focus on optimization of the indexing overhead, e.g., through computing
the integer division magic numbers on the host prior to launching the
coarse-grid kernel~\cite{int_fastdiv}.

In deploying the coarse-grid operator on multiple nodes, halo packing
and exchange routines are needed as well.  To achieve good bandwidth
saturation on the packing kernel, a fine-grained parallelization
strategy over site, color and spin was deployed.  Given that halo
exchange is \(O(\hat{N}_s\hat{N}_c)\)
but stencil application is \(O(\hat{N}_s^2\hat{N}_c^2)\)
the halo-exchange cost is relatively negligible.  Thus we focussed on
minimizing latency of the communication: a single packing kernel is
used for all exchange dimensions followed by a single copy to the CPU
of the resulting buffer.  MPI is used to exchange the halos and a
subsequent single copy to the GPU is utilized.  In this initial
implementation overlapping of computation and communication is not
performed (unlike on the fine grid which is optimized for throughput).

\vspace{-1mm}
\subsection{Inter-grid Transfer Operators}
\vspace{-1mm}
\label{sec:fine-transfer}

The transfer operators (prolongator and restrictor) are special since
they involve both fine and coarse grids.  To maximize parallelism we
always parallelize with respect to the fine-grid geometry.  For the
prolongator this is trivial, since the output is on the fine grid, so
it can be implemented as a gather operation, and we can easily assign
independent threads to each fine-grid degree of freedom (grid, color,
spin).  The restrictor is not so simple, since here the input is a
fine-grid field, and trivial parallelism over the fine-grid degrees of
freedom would imply a scatter operation (onto the coarse grid),
requiring atomics to avoid a race between threads.  The solution
is twofold: we parallelize over the output (coarse) color and spin, and
parallelize over the input grid geometry.  To avoid a race condition
on the latter we ensure that a single aggregate (for a given coarse
color and spin) is always mapped exactly to a thread block: each
thread then performs the necessary rotation from fine degrees of
freedom (color and spin) to coarse degrees of freedom, and then a
shared-memory reduction obtains the final coarse-grid result, with
the first thread in the thread block writing out the final coarse-grid
result.  For rapid development, as well as high performance, we
utilized the cub C++ header library for this~\cite{cub}.

\vspace{-1mm}
\subsection{Summary}
\vspace{-1mm}

In this section we have described how the many sources of parallelism
in a stencil problem can be exposed and mapped to the GPU architecture.
Critical in achieving this mapping in a flexible fashion is the scalar
SIMT programming model that GPUs feature, utilizing a hierarchical
Cartesian grid for expressing locality, as well as the low-latency
shared-memory that clusters of threads can utilize as a common
scratchpad to share data.  These features enable us to extract enough
parallelism to saturate the GPU on all but the smallest problem
sizes.

Due to the achieved high saturation of GPU performance, we have not
pursued an optimized CPU coarse-grid implementation
beyond rudimentary OpenMP parallelization.  We discuss our long-term
expectations with respect to CPU versus GPU for the coarse-grid
computation in \S\ref{sec:future-work}.

\section{Results}
\label{sec:results}

\subsection{Methodology}
\label{sec:methodology}
All the results for this section were carried out on the Titan System,
hosted at Oak Ridge Leadership Computing Facility (OLCF).  We used a
development version of QUDA and Chroma. The codes were
compiled using the GNU C/C++ compiler (gcc) v4.9.0, and CUDA-toolkit
version 7.0 as installed on Titan.  We carried out solves with the
MG solver in QUDA, and compared performance with QUDA's
mixed-precision BiCGStab algorithm.

We selected three exemplar
gauge configurations for this study that are representative of present
LQCD computations on Leadership facilities, the parameters of which we
list in Table \ref{tab:lattices}: $L_s$ and $L_t$ refer to the number of lattice sites in the 3 spatial and 
the temporal direction respectively; the lattice spacings $a_s$ and $a_t$ refer
to the lattice spacings in femtometers, $m_q$ is the bare sea quark mass parameter 
used for the light quarks in the configuration generation, and $m_{\pi}$ gives the mass of the 
$\pi$-meson measured on the lattices to indicate the lightness of the sea quarks. 

In our studies
the solves were carried out with the quark mass parameter equal to the sea quark masses
on these configurations.

\begin{table}
\caption{Lattice configurations and their physical parameters}
\label{tab:lattices}
\centering
\begin{tabular}{cccccccc}
\hline
Label & $L_s$ & $L_t$ & $a_s$(fm) & $a_t$ (fm) & $m_q$ & $m_{\pi}$ (MeV)  \\
\hline 
Aniso40 & $40$ & $256$ & $0.125$ & $0.035$ & $-0.0860$ & $\approx 230$  \\
Iso48     & $48$ &  $ 96$ & $0.075$& $0.075$ & $-0.2416$ &  $\approx 192$  \\
Iso64     & $64$ &  $128$ & $0.075$ & $0.075$ & $-0.2416$ & $\approx 192$  \\
\hline
\end{tabular}
\end{table}

On each configuration we computed a ``propagator'', which consists of
12 independent linear solves, and compared the time to solution
between MG and BiCGStab.  Since in the first solve QUDA
performs performance autotuning, the wallclock time for these was
artificially long, and so in our results we have averaged the
wallclock times on the last 11 solves following this.  For computing
the lattice-averaged quantities (e.g., residual) we use the \(L_2\)
norm, and estimate the solver error using the double-solve strategy
advocated in~\cite{Osborn:2010mb}.   For 
speedups we take the ratio of wallclock times and averaging the
ratios: in other words, the solves are treated as though they are
correlated.  We do not include the MG set-up time because in a
throughput calculation this time is completely amortized by a very
large number of solves. For example in hadron spectroscopy
calculations $O(10^5)-O(10^6)$ solves may be carried out per gauge
configuration. Optimizing the setup both algorithmically and in terms
of implementation is beyond the scope of our discussion here but will
be covered in future work.

For MG, we used a three-level K-cycle: on the fine and
intermediate levels we used a recursively preconditioned general
conjugate residual (GCR) algorithm with a Krylov subspace size of 10
vectors.  For smoothing on these levels four pre and post applications
of minimal residual (MR) was used and the coarse-grid solver was
GCR.  GCR was chosen as the outer solver since it is a flexible solver, thus is
tolerant of the variable precondioner that arises from using an
MR smoother in a K-cycle.  On all levels we utilized red-black preconditioning.  On the
finest level, in the MR smoother we used half precision, and we
compressed the gauge field to eight real numbers.  With the exception
of using double precision on the outer most GCR solver, all other
computation was in single precision.  The BiCGStab algorithm employed
half precision with double-precision reliable updates and
similarly utilized gauge compression.

In our tests, we investigated three MG sub-space size combinations: a)
24 null vectors on both level 1 and level 2, b) 24 null vectors on
level 1, 32 null vectors on level 2 and c) 32 null vectors on both
level 1 and level 2.  These representing typical choices in a CPU MG
method.  We refer to these strategies as $24/24$, $24/32$ and $32/32$
respectively.

Given our interest in the analysis phase of LQCD for the 
Aniso40 and Iso48 data we have only investigated small partiton sizes of 20 and 32 nodes for 
Aniso40 and 24 and 48 nodes for Iso48 respectively (note for {\em Aniso40}, the $32/32$ case did not fit on 20
nodes). For the Iso64 dataset we found the minimum
partition size to be 64 nodes, and since the lattices are relatively large, we scaled this out to 512 nodes.
Our current implementation cannot scale beyond this node count, since in this case the size of the 
coarsest lattice is $2^4$ sites per node, which is the minimum our implementation can handle. 

\begin{table}
\caption{Chief parameters for the MG tests.}
\label{tab:solveparameters}
\centering
\begin{tabular}{ccccc} \\
\hline
Label & Nodes & Level 1 & Level 2  & target \\
          &            &  blocking & blocking & residuum \\
\hline
Aniso40 & $20$ &  $5^2\times2\times 8$ & $2^3 \times 4$  & $5 \times 10^{-6}$ \\
Aniso40 & $32$ & $5^3\times8$ & $2^3\times4$ & $5 \times 10^{-6}$ \\
Iso 48 &  $24,48$ & $4^4$ & $3^3 \times 2 $ & $10^{-7}$ \\
Iso 64 & $64,128,256,512$ & $4^4$ & $2^4$ & $10^{-7}$ \\
\hline
\end{tabular}
\end{table}

\subsection{Scaling Results}

\begin{table*}[t]

\centering
\caption{Results Table.  Mean values are shown with standard deviation in
brackets.}
\leavevmode
\tiny
\label{tab:results}
\centering
\begin{tabular}{cc|cccc|cccccc}
\hline
& & \multicolumn{4}{c}{BiCGStab} & \multicolumn{6}{c}{Multigrid} \\
Label & Nodes & Iter. & Time(s) & Error/ & Nodes &
    Strategy & Iter. & Time(s) & Error/ & Nodes & Speedup\\
& & &  & Residual & \(\times\) Time&
&  & & Residual & \(\times\) Time& \\
\hline 
\multirow{5}{*}{Aniso40} &  \multirow{2}{*}{20} & \multirow{2}{*}{1771
                                                  (86)} &
                                                     \multirow{2}{*}{22.6
                                                     (1.9)} & \multirow{2}{*}{137 (38)} & \multirow{2}{*}{452} &
                                                                      24/24
                                        & 15.3 (0.5) & 2.9 (0.1) &
                                                                   42.9
                                                                   (2.2)
             & 58.0 & 7.7 (0.6) \\ 
 &  &  &  &  & &  24/32 & 14.2 (0.4) & 2.9 (0.1) & 30.2 (1.2)& 58.0 & 7.9 (0.7) \\[0.075cm]
 &  \multirow{3}{*}{32} & \multirow{3}{*}{1817 (139)} &
                                                        \multirow{3}{*}{11.8 (0.9)} & \multirow{3}{*}{134 (42)} & \multirow{3}{*}{338} &
                                                                   24/24
                                        &  17.6 (0.5) & 2.01 (0.04) &
                                                                      36.6
                                                                      (7.2)
             & 64.3 & 5.5 (1.2)\\
 &  &  &  &  &  &  24/32 & 17.9 (0.3) & 1.95 (0.07) & 43.8 (2.2) & 62.4 & 6.0 (0.5) \\ 
  &  &  &  &  &  &  32/32&  14.0 (0.0) &  2.09 (0.03) & 26.1 (1.2) &
                                                                     66.9 & 5.6 (0.5)\\  \hline 

\multirow{6}{*}{Iso48} &  \multirow{3}{*}{24} & \multirow{3}{*}{3402 (132)} &
                                                   \multirow{3}{*}{20.4
                                                   (1.3)} &
                                                            \multirow{3}{*}{110 (33)} & \multirow{3}{*}{490} & 24/24
                                        & 17.4 (0.5) & 3.84 (0.13) &
                                                                     24.9
                                                                     (1.8)
             &92.2 & 5.3 (0.2)
                                                                   \\ 
 &  &  &  &  &  &  24/32 & 17.3 (0.5) & 3.12 (0.10) & 26.8 (4.5) &
                                                                   74.9 & 6.6 (0.5)\\
 &  &  &  &  &  &  32/32 &  14.0 (0.0) & 4.16 (0.13) & 25.1 (4.5) &
                                                                    99.8 & 5.1 (0.4)\\[0.075cm]

 &  \multirow{3}{*}{48} &  \multirow{3}{*}{3522 (245)}  &  \multirow{3}{*}{14.4 (1.0)} &
                                                            \multirow{3}{*}{99.8
                                                                                         (29.2)} & \multirow{3}{*}{691}&
                                                                  24/24
                                &  17.2 (0.4) & 2.23 (0.05) & 25.6
                                                              (2.1)&
                                                                     107 & 6.3 (0.4) \\
 &  &  &  &  &  &  24/32 & 17.0 (0.0)  &  2.36 (0.07) & 25.1 (2.0)& 113& 6.1 (0.4)\\ 
  &  &  &  &  &  &  32/32&  14.0 (0.0) & 2.84 (0.07) & 25.9 (2.0)& 136& 5.1 (0.4)\\  \hline 

\multirow{12}{*}{Iso64} &  \multirow{3}{*}{64} &  \multirow{3}{*}{2805
                                                 (159)} & \multirow{3}{*}{22.2 (1.7)} & 
                                                                     \multirow{3}{*}{210 (84)} &\multirow{3}{*}{1421} &
                                                                     24/24
                                        & 17.4 (0.5) & 4.11 (0.15) & 
                                                                      29.9
                                                                     (2.3)
  & 263 & 5.4 (0.4)\\ 
 &  &  &  &  &  &  24/32 & 17.0 (0.0)  & 4.48 (0.96)  & 25.7 (1.7) &
                                                                     287 &
  5.1 (0.8)\\
 &  &  &  &  &  &  32/32 &  14.0 (0.0) & 4.63 (0.15) & 31.4 (7.4) &296 &
                                                                      4.8 (0.3)\\[0.075cm]
 &  \multirow{3}{*}{128} &  \multirow{3}{*}{2807 (171)} &
                                                          \multirow{3}{*}{30.7 (2.4)}  &
                                                                   \multirow{3}{*}{199 (90)} &\multirow{3}{*}{3930} &
                                                                   24/24
                                        & 18.0 (0.0) & 3.01 (0.06) &
                                                                     33.6
                                                                     (1.5)
             &385 & 10.2 (0.7) \\
 &  &  &  &  &  &  24/32 &  16.7 (0.5) & 3.05 (0.07) & 24.7 (1.8) &
                                                                    390 &
                                                                    10.1 (0.6)\\ 
  &  &  &  &  &  &  32/32&  14.0 (0.0) & 3.46 (0.05) & 31.8 (9.3) &443
                               & 8.9 (0.6)\\[0.075cm]
&  \multirow{3}{*}{256} & \multirow{3}{*}{2885 (171)} &
                                                        \multirow{3}{*}{22.5 (1.8)}
              & \multirow{3}{*}{191 (76)} & \multirow{3}{*}{5760} &
                                                                    24/24
                                        & 18.0 (0.0) & 2.36 (0.07) &
                                                                     32.0
                                                                     (4.1)
             & 604 & 9.5 (0.8)\\ 
 &  &  &  &  &  &  24/32 &  16.4 (0.5) & 2.12 (0.08) & 24.5 (2.0) & 543&
                               10.6 (0.8)\\
 &  &  &  &  &  &  32/32 &  14.0 (0.0) & 2.37 (0.06) & 32.1 (5.3) &
                                                                    607 &
                                                                      9.5
  (0.7)\\[0.075cm]
 &  \multirow{3}{*}{512} & \multirow{3}{*}{2940 (269)} &
                                                         \multirow{3}{*}{
                                                         12.3 (0.7)}
      & \multirow{3}{*}{198 (80)} & \multirow{3}{*}{6298} &
                                                                     24/24
                                        & 17.9 (0.3) & 1.73 (0.08)  &
                                                                      33.2
                                                                      (2.0)
             & 886 & 7.1 (0.4)\\
 &  &  &  &  &  &  24/32 &  17.0 (0.0) & 1.97 (0.10) & 25.8 (2.0) & 1009& 6.3 (0.3)\\ 
  &  &  &  &  &  &  32/32 &  13.7 (0.5) & 1.93 (0.13) & 33.4 (5.8) & 988&
  6.4 (0.2)\\
\hline
\end{tabular}
\end{table*}

\begin{figure*}[t]
\centering
\includegraphics[trim={0 1.5cm 0 7.5cm},clip,width=5in]{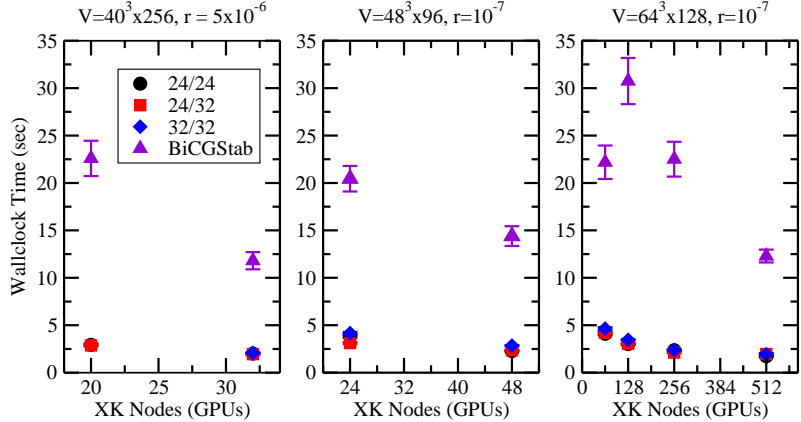}%
\vspace{-2mm}
\caption{MG scaling vs BiCGStab for the datasets given in 
  Table \ref{tab:lattices}, algorithm parameters given in 
  \ref{tab:solveparameters} and 
  \S\ref{sec:methodology}. \label{fig:scalingbicg}}
\end{figure*}

Our results are given in Table \ref{tab:results}, where we
compare the iteration count, average wallclock time, error/residual
ratio and cost (defined as number of nodes \(\times\)
time) for the solvers, as well as the MG wallclock speedup
versus BiCGStab at constant node count.  Figure
\ref{fig:scalingbicg} is a visual repsentation of these 
wallclock times.  One of the propagator solves on the $24/24$ case took
anomalously long: we have excluded this, together with the
corresponding solves from the $24/32$, $32/32$ and BiCGStab
results. Likewise, we found an anomalously large value for MG with
$24/32$ for the {\em Iso64} workload; we again treated it as an
anomaly and excluded that component from the analysis for
all the solver setups. These exclusions did not change the average
values significantly, but rather changed the error bars which were
otherwise excessively large.  Our experience has been that such
fluctations are commonplace when running on Titan and we hypothesis are due to
cross-job network polution.

In going from 64 nodes to 128 nodes the BiCGStab runtime increases,
rather than decreases, though for partitions larger than 128 nodes
BiCGStab runtime scales reasonably. This observation was repeatable
and does not occur with MG.  We attribute this to node placement for
the job, possibly due to having the job fit into one cabinet on 64
nodes, to no longer fitting into one cabinet for 128 nodes.  Moreover, 
BiCGStab is more strictly communications limited compared to MG's
more latency-limited profile.  

The speedup of MG over BiCGStab is
typically between 5--8\(\times\),
although it exceeds \(10\times\)
in the case of the {\em Iso64} dataset on 128 and 256 nodes.  In all
cases the minimum cost occurs on the least numbers of nodes,
demontrating the difficulty in strong scaling sparse iterative
solvers.  In most
cases the 24/24 and 24/32 strategies proved to be superior; while
32/32 gives a better preconditioner since it
captures more of the null space, the increased cost of the
intermediate grid results in a net computational loss. 

The drastically reduced and stable iteration count of MG demonstrates its numerical
robustness compared to the more chaotic convergence of
BiCGStab.  The reduced error / residual ratio is indicative
of MG's ability to damp both the high and low error modes uniformly: thus the
cost reduction at constant error will be even greater than the
na\"{i}ve speedups shown in Table \ref{tab:results}. 

We do not focus on raw GFLOPS rates here, but we note that MG
typically sustains around 3--5\(\times\)
less GFLOPS than BiCGStab when running on the same number of GPUs.
Using the {\tt nvidia-smi} utility we compared the power efficiency of
the two algorithms: typically MG
consumes \(\sim15\%\)
less power than BiCGStab, e.g., for {\it Iso48} on 48 nodes, node 0 used 72W
 for MG versus 83W for BiCGStab. This is not surprising given the
reduced GFLOPS performance of MG versus BiCGStab; hence MG is
also more power efficient as well as more time efficient.

\begin{figure}[htb]
\centering
\includegraphics[trim={0.1cm 0.3cm 0.1cm 0.3cm},clip,width=3.1in]{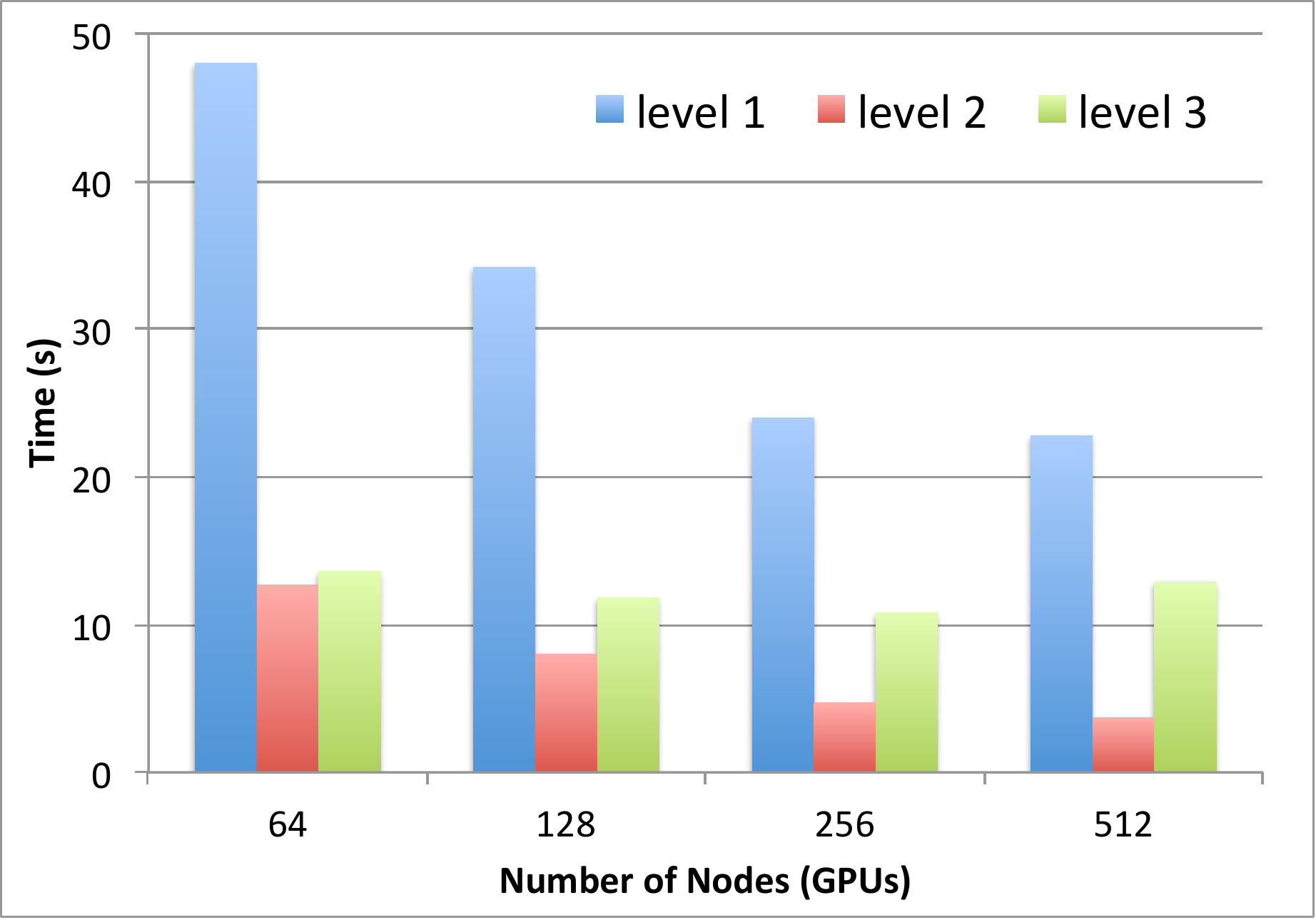}%
\vspace{-2mm}
\caption{Breakdown of time spent in the different MG levels
  for the {\it Iso64} dataset with the 24/32 strategy.\label{fig:breakdown}}
\end{figure}

To better understand the rate-limiting step of the MG solver, in
Figure \ref{fig:breakdown} we show the breakdown of time spent in
three different levels of the solver for the large {\em Iso64} problem as a
function of the number of nodes.  It can be seen that the coarsest grid
constitutes an ever increasing fraction of the time spent.  Closer
profiling revealed this to be due to the global synchronizations
present in the coarse-grid GCR solver: the \( \log N\) scaling of
the cost of synchronization dominates that of the stencil application
at large node count.

\section{Software Implications of Fine-grained Parallelization}
\label{sec:software}

As noted in \S\ref{sec:fine}, the most commonly deployed approach to
parallelizing grid-based stencil computation is to employ the data
parallelism of the underlying grid.  In LQCD this approach is
particularly attractive, since it means that the underlying physics can
be expressed in terms of the natural matrices, vectors, etc. that
underpin these theories.  It is exactly this strategy that
domain-specific languages (DSLs) have employed in LQCD: large code
bases have been built on this principle, e.g., Chroma, the newly
introduced Grid framework~\cite{Boyle:2015tjk}, etc.  With this
data-parallel abstraction, it is then possible to deploy these
applications on a variety of architectures in a portable fashion.
However, by expressing computation at the grid level, this excludes
the kind of fine-grained parallelization strategies employed in this
work.  With a hierarchy of scales, MG certainly exaggerates the need
for this in today's architecture, however, as we move to ever more
parallel systems, the need for fine-grained parallelism will
eventually extend in a more intrusive manner into software, e.g., not
just in the coarse-grid of a MG solve.  Given potential super-linear
scaling of algorithmic cost with respect to problem size, we cannot
rely upon weak scaling in order to saturate future machines and meet
the expected science fidelity goals.  In planning for the scalable
software models for the Exascale and beyond, we advocate a
multi-resolution approach for future frameworks, by which we mean
interfaces to work both at the level of grid parallelism or at levels
where the extra degrees of parallelism are fully exposed.  The
majority of the code can then be written in the simpler style as
before, but frameworks would be able to also accommodate the
parallelism needed for performance critical sections. This statement is
not specific to GPUs, e.g., the recent Intel Knights Landing
architecture with up to 72 cores, each with wide vectors units and multiple
SIMT threads, may well require similar optimization strategies.

\section{Future Work}
\label{sec:future-work}
This work focussed on accelerating throughput workloads, but not on
the strong scaling workloads.  Future work will focus on this area,
in particular through the use of Schwarz-style communication-reducing
preconditioners to improve strong scaling of the MG
smoothers~\cite{Frommer:2013kla,Frommer:2013fsa}, and
through replacement of the coarse-grid solver with a latency
tolerant solver, such as CA-GMRES~\cite{Hoemmen:2010:CKS:1970638,6877343}.

Another avenue to increase parallelism is to reformulate MG as a
multiple-right-hand-side solver, where we solve multiple linear
systems simultaneously.  For \(N\)
right hand sides, we thus expose \(N-\)way
additional parallelism, as well as increasing the temporal locality of
the problem, e.g., the same stencil operator is used for all systems.
We see this as another critical reformulation to maintain scalability
on future architectures (c.f. \cite{Richtmann:2016kcq}).

While at present, GPUs are efficient for the entire MG computation, we
expect in the future coarse grids will favor latency-optimized
CPUs since throughput-optimized processors (e.g., GPUs) will continue to
get wider and at some point exhaust all available
parallelism.  When this point is reached, as it has been in other MG
formulations, e.g.,~\cite{hpgmg-gpu}, the algorithm will then be
optimally deployed on the heterogeneous architecture.  It is then
tempting to wonder whether we can utilize the CPU and GPU
simultaneously using an {\it additive} correction at the boundary
between architectures.  While additive multigrid is known not to
extend to multiple levels, in this case we only seek an additive
correction at the interface, and maintain a multiplicative correction
elsewhere.

\section{Conclusions}
\label{sec:conclusion}

This report is a step in a critical transition for LQCD, as it
explores finer resolution with multiple physical scales and adapts to
increasingly parallel architectures on the path to the Exascale.  MG is
a dramatic example of the huge algorithmic potential of incorporating
multiple scales in software infrastructure. Algorithmic gains of
one to two orders of magnitude are too compelling and must be
accommodated.  However, the hyper parallelism of many-core
architectures often conflicts with the multi-scale algorithmic needs
that require fine-grain parallelism and non-local memory
references. We believe this article is a prototypical example of the
substantial software restructuring needed to combine multi-scale
algorithms with future architectures across many disciplines of
computational science.

Already precision results in lattice LQCD are competing in accuracy
with the best experimental results. This in turn has established LQCD
as a necessary partner with experiment in the fundamental search for
new physics beyond the Standard Model at the Large Hardon Collider and
at nuclear and particle physics laboratories around the
world. For LQCD to continue to advance in this enterprise, further
research along the lines in this limited study must proceed in the
next decade to maintain the momentum in the exploration of particle
and nuclear physics.

\section*{Acknowledgments}

The authors would like to thank Chip Watson, for hosting a GPU
Multigrid Hackathon at Jefferson Lab, and Judy Hill and Don Maxwell
of Oak Ridge Leadership Computing Facility (OLCF) for assistance with
scripting the power measurements on Titan. This authors thank the
USQCD collaboration for devoting a portion of their OLCF INCITE
allocation LGT003 for measurements and benchmarks for this work, and
for providing the test gauge configurations which were generated under
OLCF INCITE, ALCC and NSF PRAC allocations. B.~Jo\'o gratefully
acknowledges funding by the U.S. Department of Energy, Office of
Science, Office of Advanced Scientific Computing Research and Office
of Nuclear Physics under the U.S. D.O.E. Scientific Discovery through
Advanced Computing (SciDAC) program. This material is based upon work
supported by the U.S. Department of Energy, Office of Science, Office
of Nuclear Physics under contract DE-AC05-06OR23177. Work by A.S. was
done for Fermi Research Alliance, LLC under Contract No.
DE-AC02-07CH11359 with the United States Department of Energy.
This research used resources of the Oak Ridge Leadership Computing Facility at the Oak Ridge National Laboratory, which is supported by the Office of Science of the U.S. Department of Energy under Contract No. DE-AC05-00OR22725.

\section*{Appendix: Reproducibility}
The source code of the programs used, the input and output data,
run-scripts and post processing scripts used to generate the data
in this paper can be made available on request, along with assistance
in code-building, installation and running. Please contact 
B. Jo\'o for details.

\bibliographystyle{abbrv}
\bibliography{sc16}
\end{document}